\documentclass[aps,prl,twocolumn,superscriptaddress,floatfix,nofootinbib,showpacs,amsmath,amssymb,altaffilletter]{revtex4-1}

\usepackage{graphicx}% Include figure files
\usepackage{dcolumn}% Align table columns on decimal point
\usepackage{bm}% Bold math
\usepackage{color}% Colored text
\usepackage{verbatim}% Allows long comments
\usepackage{paralist}% Compact environments
\usepackage{lineno}% Line numbers
\usepackage{subfigure}

\newcommand{\ber}{\mbox{$^{7}$Be}}
\newcommand{\ncno}{\mbox{$^{13}$N}}
\newcommand{\ocno}{\mbox{$^{15}$O}}
\newcommand{\fcno}{\mbox{$^{17}$F}}
\newcommand{\pep}{\mbox{\it pep}}
\newcommand{\pp}{\mbox{\it pp}}
\newcommand{\CNO}{\mbox{CNO}}
\newcommand{\bor}{\mbox{$^{8}$B}}
\newcommand{\ctwe}{\mbox{$^{12}$C}}
\newcommand{\cele}{\mbox{$^{11}$C}}
\newcommand{\cten}{\mbox{$^{10}$C}}

\newcommand{\pota}{\mbox{$^{40}$K}}
\newcommand{\kr}{\mbox{$^{85}$Kr}}

\newcommand{\bite}{\mbox{$^{210}$Bi}}

\newcommand{\pbfo}{\mbox{$^{214}$Pb}}
\newcommand{\Bipo}{\mbox{$^{214}$Bi-$^{214}$Po}}

\newcommand{\tho}{\mbox{$^{232}$Th}}
\newcommand{\ura}{\mbox{$^{238}$U}}

\newcommand{\Pee}{\mbox{$P_{ee}$}}

\newcommand{\cpd}{\mbox{counts/(day$\cdot$100\,ton)}}
\newcommand{\flx}{\mbox{cm$^{-2}$s$^{-1}$}}
\newcommand{\powero}[1]{\mbox{10$^{#1}$}}
\newcommand{\powert}[2]{\mbox{#2$\times$10$^{#1}$}}

\newcommand{\tleight}{\mbox{$^{208}$Tl}}
\newcommand{\bifo}{\mbox{$^{214}$Bi}}
\newcommand{\hesix}{\mbox{$^{6}$He}}
\newcommand{\pam}{\mbox{$^{234m}$Pa}}

\begin{document}
\title{First evidence of \pep\ solar neutrinos by direct detection in Borexino}

\newcommand{\APC}{Laboratoire AstroParticule et Cosmologie, 75205 Paris cedex 13, France}
\newcommand{\Budapest}{KFKI-RMKI, 1121 Budapest, Hungary}
\newcommand{\Dubna}{Joint Institute for Nuclear Research, 141980 Dubna, Russia}
\newcommand{\Genova}{Dipartimento di Fisica, Universit\`a e INFN, Genova 16146, Italy}
\newcommand{\Hamburg}{Institut f\"ur Experimentalphysik, Universit\"at, 22761 Hamburg, Germany}
\newcommand{\Heidelberg}{Max-Planck-Institut f\"ur Kernphysik, 69117 Heidelberg, Germany}
\newcommand{\Kiev}{Kiev Institute for Nuclear Research, 06380 Kiev, Ukraine}
\newcommand{\Krakow}{M.~Smoluchowski Institute of Physics, Jagiellonian University, 30059 Krakow, Poland}
\newcommand{\Kurchatov}{NRC Kurchatov Institute, 123182 Moscow, Russia}
\newcommand{\LNGS}{INFN Laboratori Nazionali del Gran Sasso, SS 17 bis Km 18+910, 67010 Assergi, Italy}
\newcommand{\Milano}{Dipartimento di Fisica, Universit\`a degli Studi e INFN, 20133 Milano, Italy}
\newcommand{\Munich}{Physik Department, Technische Universit{\"a}t M{\"u}nchen, 85747 Garching, Germany}
\newcommand{\Pavia}{INFN, Pavia 27100, Italy}
\newcommand{\Perugia}{Dipartimento di Chimica, Universit\`a e INFN, 06123 Perugia, Italy}
\newcommand{\Peters}{St. Petersburg Nuclear Physics Institute, 188350 Gatchina, Russia}
\newcommand{\Princeton}{Physics Department, Princeton University, Princeton, NJ 08544, USA}
\newcommand{\PrincetonChemEng}{Chemical Engineering Department, Princeton University, Princeton, NJ 08544, USA}
\newcommand{\Queens}{Physics Department, Queen's University, Kingston ON K7L 3N6, Canada}
\newcommand{\UMass}{Physics Department, University of Massachusetts, Amherst, MA 01003, USA}
\newcommand{\Virginia}{Physics Department, Virginia Polytechnic Institute and State University, Blacksburg, VA 24061, USA}
\newcommand{\Valencia}{Instituto de F\`isica Corpuscular, CSIC-UVEG, Valencia 46071, Spain}

\author{G.~Bellini}\affiliation{\Milano}
\author{J.~Benziger}\affiliation{\PrincetonChemEng}
\author{D.~Bick}\affiliation{\Hamburg}
\author{S.~Bonetti}\affiliation{\Milano}
\author{G.~Bonfini}\affiliation{\LNGS}
\author{D.~Bravo}\affiliation{\Virginia}
\author{M.~Buizza~Avanzini}\affiliation{\Milano}
\author{B.~Caccianiga}\affiliation{\Milano}
\author{L.~Cadonati}\affiliation{\UMass}
\author{F.~Calaprice}\affiliation{\Princeton}
\author{C.~Carraro}\affiliation{\Genova}
\author{P.~Cavalcante}\affiliation{\LNGS}
\author{A.~Chavarria}\affiliation{\Princeton}
\author{D.~D'Angelo}\affiliation{\Milano}
\author{S.~Davini}\affiliation{\Genova}
\author{A.~Derbin}\affiliation{\Peters}
\author{A.~Etenko}\affiliation{\Kurchatov}
\author{K.~Fomenko}\affiliation{\Dubna}\affiliation{\LNGS}
\author{D.~Franco}\affiliation{\APC}
\author{C.~Galbiati}\affiliation{\Princeton}
\author{S.~Gazzana}\affiliation{\LNGS}
\author{C.~Ghiano}\affiliation{\LNGS}
\author{M.~Giammarchi}\affiliation{\Milano}
\author{M.~Goeger-Neff}\affiliation{\Munich}
\author{A.~Goretti}\affiliation{\Princeton}
\author{L.~Grandi}\affiliation{\Princeton}
\author{E.~Guardincerri}\affiliation{\Genova}
\author{S.~Hardy}\affiliation{\Virginia}
\author{Aldo~Ianni}\affiliation{\LNGS}
\author{Andrea~Ianni}\affiliation{\Princeton}
\author{D.~Korablev}\affiliation{\Dubna}
\author{G.~Korga}\affiliation{\LNGS}
\author{Y.~Koshio}\affiliation{\LNGS}
\author{D.~Kryn}\affiliation{\APC}
\author{M.~Laubenstein}\affiliation{\LNGS}
\author{T.~Lewke}\affiliation{\Munich}
\author{E.~Litvinovich}\affiliation{\Kurchatov}
\author{B.~Loer}\affiliation{\Princeton}
\author{F.~Lombardi}\affiliation{\LNGS}
\author{P.~Lombardi}\affiliation{\Milano}
\author{L.~Ludhova}\affiliation{\Milano}
\author{I.~Machulin}\affiliation{\Kurchatov}
\author{S.~Manecki}\affiliation{\Virginia}
\author{W.~Maneschg}\affiliation{\Heidelberg}
\author{G.~Manuzio}\affiliation{\Genova}
\author{Q.~Meindl}\affiliation{\Munich}
\author{E.~Meroni}\affiliation{\Milano}
\author{L.~Miramonti}\affiliation{\Milano}
\author{M.~Misiaszek}\affiliation{\Krakow}\affiliation{\LNGS}
\author{D.~Montanari}\affiliation{\LNGS}\affiliation{\Princeton}
\author{P.~Mosteiro}\affiliation{\Princeton}
\author{V.~Muratova}\affiliation{\Peters}
\author{L.~Oberauer}\affiliation{\Munich}
\author{M.~Obolensky}\affiliation{\APC}
\author{F.~Ortica}\affiliation{\Perugia}
\author{K.~Otis}\affiliation{\UMass}
\author{M.~Pallavicini}\affiliation{\Genova}
\author{L.~Papp}\affiliation{\Virginia}
\author{L.~Perasso}\affiliation{\Milano}
\author{S.~Perasso}\affiliation{\Genova}
\author{A.~Pocar}\affiliation{\UMass}
\author{J.~Quirk}\affiliation{\UMass}
\author{R.S.~Raghavan}\affiliation{\Virginia}
\author{G.~Ranucci}\affiliation{\Milano}
\author{A.~Razeto}\affiliation{\LNGS}
\author{A.~Re}\affiliation{\Milano}
\author{A.~Romani}\affiliation{\Perugia}
\author{A.~Sabelnikov}\affiliation{\Kurchatov}
\author{R.~Saldanha}\affiliation{\Princeton}
\author{C.~Salvo}\affiliation{\Genova}
\author{S.~Sch\"onert}\affiliation{\Munich}
\author{H.~Simgen}\affiliation{\Heidelberg}
\author{M.~Skorokhvatov}\affiliation{\Kurchatov}
\author{O.~Smirnov}\affiliation{\Dubna}
\author{A.~Sotnikov}\affiliation{\Dubna}
\author{S.~Sukhotin}\affiliation{\Kurchatov}
\author{Y.~Suvorov}\affiliation{\LNGS}
\author{R.~Tartaglia}\affiliation{\LNGS}
\author{G.~Testera}\affiliation{\Genova}
\author{D.~Vignaud}\affiliation{\APC}
\author{R.B.~Vogelaar}\affiliation{\Virginia}
\author{F.~von~Feilitzsch}\affiliation{\Munich}
\author{J.~Winter}\affiliation{\Munich}
\author{M.~Wojcik}\affiliation{\Krakow}
\author{A.~Wright}\affiliation{\Princeton}
\author{M.~Wurm}\affiliation{\Hamburg}
\author{J.~Xu}\affiliation{\Princeton}
\author{O.~Zaimidoroga}\affiliation{\Dubna}
\author{S.~Zavatarelli}\affiliation{\Genova}
\author{G.~Zuzel}\affiliation{\Krakow}

\collaboration{Borexino Collaboration}
\noaffiliation

\date{\today}

\begin{abstract}
We observed, for the first time, solar neutrinos in the 1.0--1.5\,MeV energy range.  We measured the rate of \pep\ solar neutrino interactions in Borexino to be (3.1$\pm$0.6$_{\rm stat}$$\pm$0.3$_{\rm syst}$)\,\cpd\ and provided a constraint on the \CNO\ solar  neutrino interaction rate of $<$7.9\,\cpd\ (95\%~C.L.).  The absence of the solar neutrino signal is disfavored at 99.97\% C.L., while the absence of the \pep\ signal is disfavored at 98\% C.L. This unprecedented sensitivity was achieved by adopting novel data analysis techniques for the rejection of cosmogenic \cele, the dominant background in the 1--2\,MeV region.  Assuming the MSW-LMA solution to solar neutrino oscillations, these values correspond to solar neutrino fluxes of \powert{8}{(1.6$\pm$0.3)}\,\flx\ and $<$\powert{8}{7.7}\,\flx\ (95\%~C.L.), respectively, in agreement with the Standard Solar Model.  These results represent the first measurement of the \pep\ neutrino flux and the strongest constraint of the \CNO\ solar neutrino flux to date.
\end{abstract}


\maketitle

Over the past 40 years solar neutrino experiments~\cite{bib:rchem-cl,bib:rchem-ga,bib:kamiokande,bib:sno,bib:bxbe7} have proven to be sensitive tools to test both astrophysical  and elementary particle physics models.  Solar neutrino detectors have demonstrated that stars are powered by nuclear fusion reactions.  Two distinct processes, the main \pp\ fusion chain  and the sub-dominant \CNO\ cycle, are expected to produce solar neutrinos with different energy spectra and fluxes.  Until now only fluxes from the \pp\ chain have been measured: \ber, \bor, and, indirectly, \pp.  Experiments involving solar neutrinos and reactor anti-neutrinos~\cite{bib:kamland} have shown that solar neutrinos undergo flavor oscillations.

Results from solar neutrino experiments are consistent with the Mikheyev-Smirnov-Wolfenstein Large Mixing Angle (MSW-LMA) model~\cite{bib:msw}, which predicts a transition from vacuum-dominated to matter-enhanced oscillations, resulting in an energy dependent $\nu_e$ survival probability, $P_{ee}$.  Non-standard neutrino interaction models formulate $P_{ee}$ curves that deviate significantly from MSW-LMA, particularly in the 1--4\,MeV transition region, see e.g.~\cite{bib:nonstandard}.  The mono-energetic 1.44\,MeV \pep\ neutrinos, which belong to the \pp\ chain and whose Standard Solar Model (SSM) predicted flux has one of the smallest uncertainties (1.2\%) due to the solar luminosity constraint \cite{bib:ssm2011}, are an ideal probe to test these competing hypotheses.

The detection of neutrinos resulting from the \CNO\ cycle has important implications in astrophysics, as it would be the first direct evidence of the nuclear process that is believed to fuel massive stars ($>$1.5$M_{\odot}$).  Furthermore, its measurement may help to resolve the solar metallicity problem~\cite{bib:metallicity, bib:ssm2011}.  The energy spectrum of neutrinos from the \CNO\ cycle is the sum of three continuous spectra with end point energies of 1.19 (\ncno), 1.73 (\ocno) and 1.74\,MeV (\fcno), close to the \pep\ neutrino energy.  The total \CNO\ flux is similar to that of the \pep\ neutrinos but its predicted value is strongly dependent on the inputs to the solar modeling, being 40\% higher in the High Metallicity (GS98) than in the Low Metallicity (AGSS09) solar model~\cite{bib:ssm2011}.

Neutrinos interact through elastic scattering with electrons in the $\sim$278 ton organic liquid scintillator target of Borexino~\cite{bib:bxdetectorpaper}.  The electron recoil energy spectrum from \pep\ neutrino interactions in Borexino is a Compton-like shoulder with end point of 1.22\,MeV.  High light yield and unprecedentedly low background levels~\cite{bib:bxbe7,bib:bxliquid} make Borexino the only detector presently capable of performing solar neutrino spectroscopy below 2\,MeV.  Its potential has already been demonstrated in the precision measurement of the 0.862\,MeV \ber\ solar neutrino flux \cite{bib:bxbe7,bib:daynight}.  The  detection of \pep\ and \CNO\ neutrinos is even more challenging, as their expected interaction rates are $\sim$10~times lower, a few counts per day in a 100\,ton target.

\begin{figure}[!t]
\centering
\includegraphics[width=\linewidth]{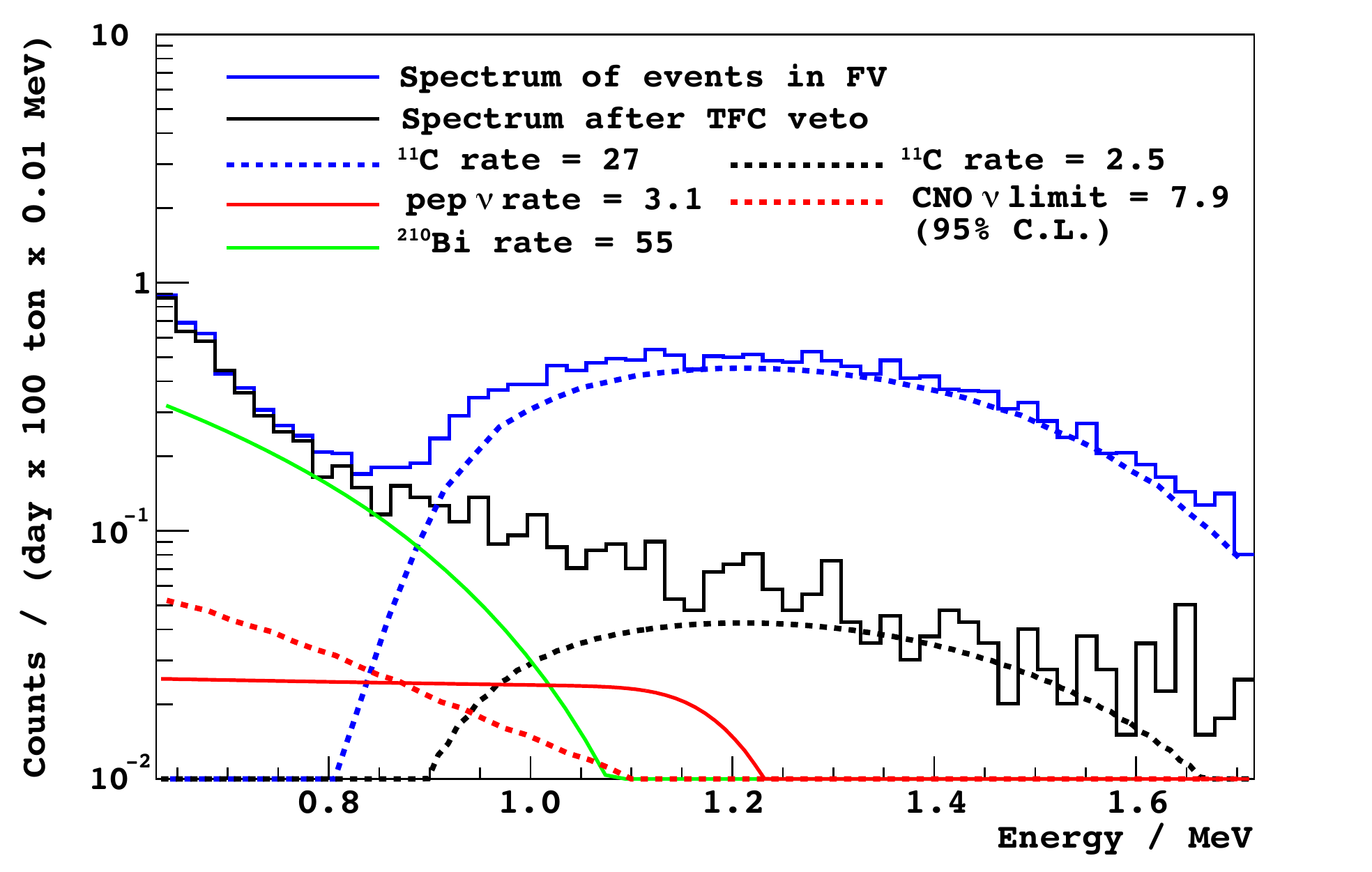}
\includegraphics[width=\linewidth]{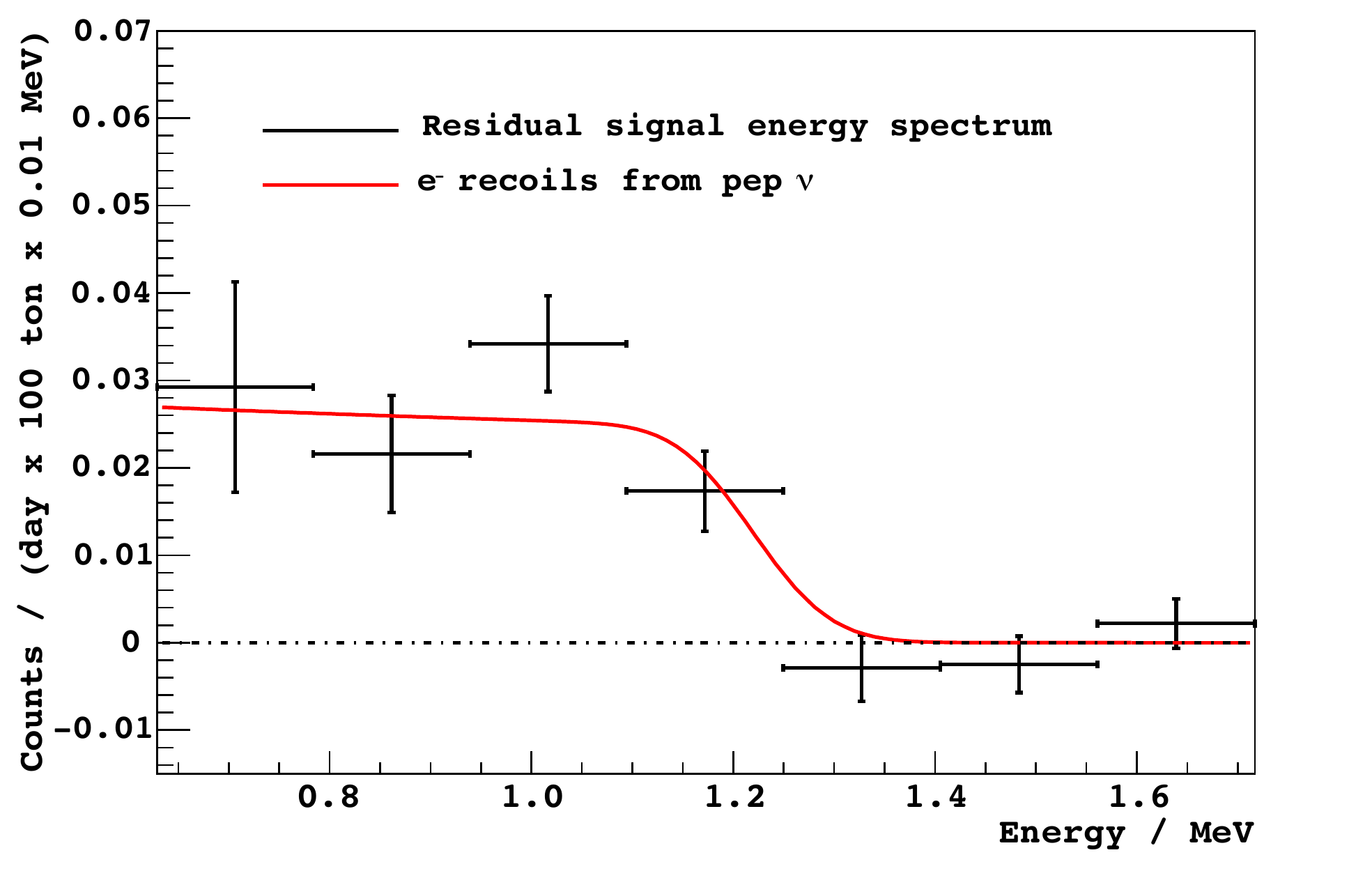}
\caption{
Top: energy spectra of the events in the FV before and after the TFC veto is applied.  The solid and dashed blue lines show the data and estimated \cele\ rate before any veto is applied.  The solid black line shows the data after the procedure, in which the \cele\ contribution (dashed) has been greatly suppressed.  The next largest background, \bite, and the electron recoil spectra of the best estimate of the \pep\ neutrino rate and of the upper limit of the \CNO\ neutrino rate are shown for reference.  Rate values in the legend are quoted in units of counts/(day$\cdot$100\,metric\,ton).
Bottom: residual energy spectrum after best-fit rates of all considered backgrounds are subtracted.  The electron recoil spectrum from \pep\ neutrinos at the best-fit rate is shown for comparison.
}
\label{fig:tfc}
\end{figure}

\begin{figure}[!t]
\centering
\includegraphics[width=\linewidth]{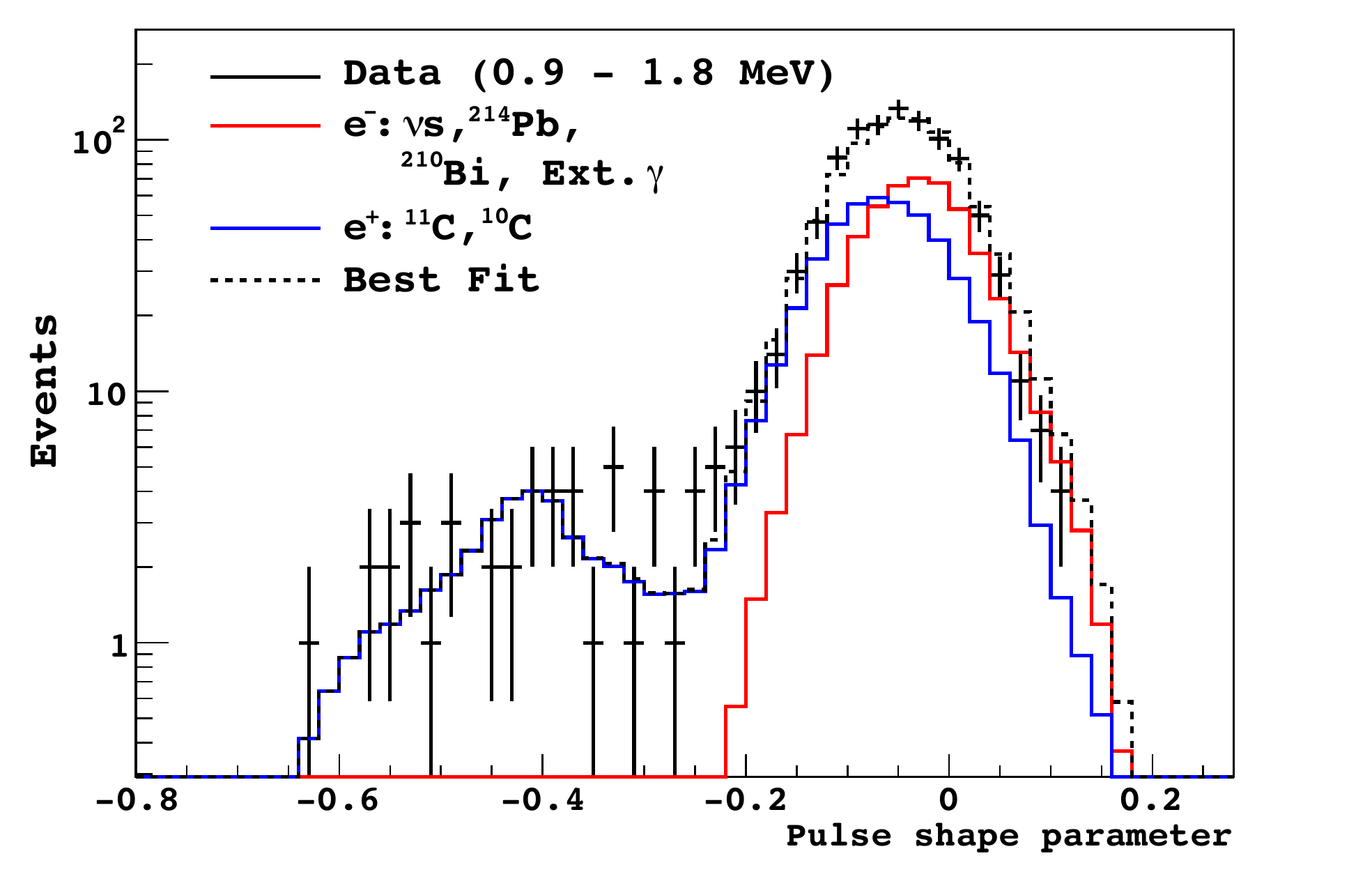}
\caption{Experimental distribution of the pulse shape parameter (black). The best-fit distribution (black dashed) and the corresponding $e^-$ (red) and $e^+$ (blue) contributions are also shown.}
\label{fig:bdt}
\end{figure}

We adopted novel analysis procedures to suppress the dominant background in the 1--2\,MeV energy range, the cosmogenic $\beta^+$-emitter \cele\ (lifetime: 29.4 min).  \cele\ is produced in the scintillator by cosmic muon interactions with \ctwe\ nuclei.  The muon flux through Borexino is $\sim$4300\,$\mu$/day, yielding a \cele\ production rate of $\sim$27\,\cpd.  In 95\%~of the cases at least one free neutron is spalled in the \cele\ production process \cite{bib:c11cris}, and then captured in the scintillator with a mean time of 255\,$\mu$s \cite{bib:bxmuon}.  The \cele\ background can be reduced by performing a space and time veto after coincidences between signals from the muons and the cosmogenic neutrons \cite{bib:deutsch, bib:pep-ctf}, discarding exposure that is more likely to contain \cele\ due to the correlation between the parent muon, the neutron and the subsequent \cele\ decay (the Three-Fold Coincidence, TFC).  The technique relies on the reconstructed track of the muon and the reconstructed position of the neutron-capture $\gamma$-ray \cite{bib:bxmuon}. The rejection criteria were chosen to obtain the optimal compromise between \cele\ rejection and preservation of fiducial exposure, resulting in a \cele\ rate of (2.5$\pm$0.3)\,\cpd, (9$\pm$1)$\%$ of the original rate, while preserving 48.5\% of the initial exposure.  The resulting spectrum (Fig.~\ref{fig:tfc}, top) corresponds to a fiducial exposure of 20409 ton$\cdot$day, consisting of data collected between January 13, 2008 and May 9, 2010.

The \cele\ surviving the TFC veto is still a significant background.  We exploited the pulse shape differences between $e^-$  and $e^+$ interactions in organic liquid scintillators \cite{bib:annihilation, bib:positronium} to discriminate \cele\ $\beta^+$ decays from neutrino-induced $e^-$ recoils and $\beta^-$decays.  A slight difference in the time distribution of the scintillation signal arises from the finite lifetime of ortho-positronium as well as from the presence of annihilation $\gamma$-rays, which present a distributed, multi-site event topology and a larger average ionization density than electron interactions.  An optimized pulse shape parameter was constructed using a boosted-decision-tree algorithm \cite{bib:tmva}, trained with a TFC-selected set of \cele\ events ($e^+$) and \bifo\ events  ($e^-$) selected by the fast \Bipo\ $\alpha$-$\beta$ decay sequence.

We present results of an analysis based on a binned likelihood multivariate fit performed on the energy, pulse shape, and spatial distributions of selected scintillation events whose reconstructed position is within the fiducial volume (FV), i.e. less than 2.8\,m from the detector center and with a vertical position relative to the detector center between -1.8\,m and 2.2\,m. We confirmed the accuracy of the modeling of the detector response function used in the fit by means of an extensive calibration campaign with $\alpha$, $\beta$, $\gamma$ and neutron sources deployed within the active target~\cite{bib:bxbe7}.

The distribution of the pulse shape parameter (Fig.~\ref{fig:bdt}) was a key element in the multivariate fit, where decays from cosmogenic \cele\ (and \cten) were considered $e^+$ and all other species $e^-$.

\begin{figure}[!t]
\centering
\includegraphics[width=\linewidth]{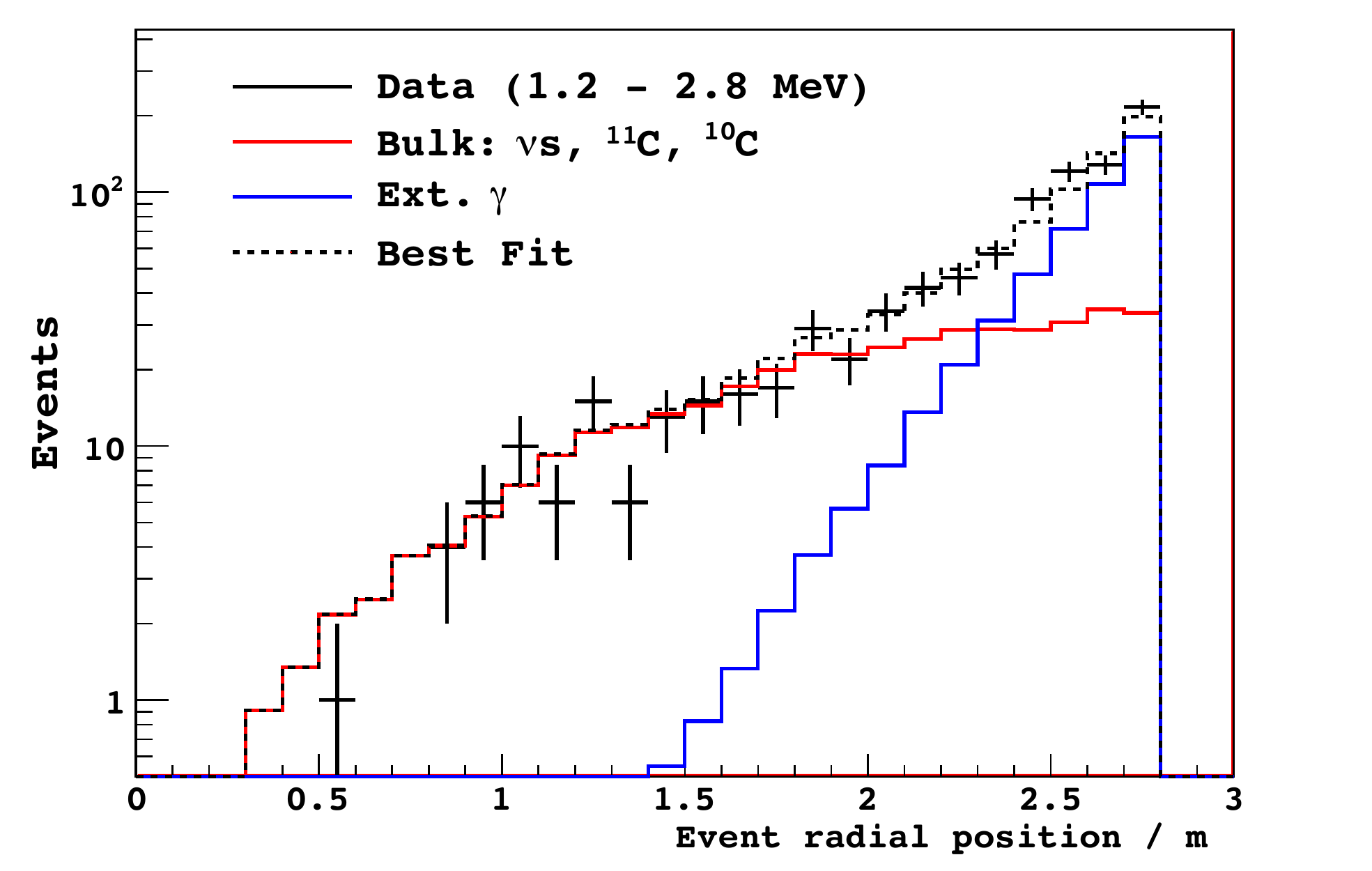}
\caption{
Experimental distribution of the radial coordinate of the reconstructed position within the FV (black).  The best-fit distribution (black dashed) and the corresponding contributions from bulk events (red) and external $\gamma$-rays (blue) are also shown.
}
\label{fig:rdist}
\end{figure}

The energy spectra and spatial distribution of the external $\gamma$-ray backgrounds have been obtained from a full, Geant4-based Monte Carlo simulation, starting with the radioactive decays of contaminants in the detector peripheral structure  and propagating the particles into the active volume.  We validated the simulation with calibration data from a high-activity $^{228}$Th source~\cite{maneschg} deployed in the outermost buffer region, outside the active volume.  The non-uniform radial distribution of the external background was included in the multivariate fit and strongly constrained its contribution.  Neutrino-induced $e^-$ recoils and internal radioactive backgrounds were assumed to be uniformly distributed. Fig.~\ref{fig:rdist} shows the radial component of the fit.

We removed $\alpha$ events  from the energy spectrum by the method of statistical subtraction~\cite{bib:bxbe7}.  We excluded from the fit all background species whose rates were estimated to be less than 5\% of the predicted rate from \pep\ neutrinos in the energy region of interest.  Furthermore, we constrained all rates to positive values.  The thirteen species left free in the fit were the internal radioactive backgrounds \bite, \cele, \cten, \hesix, \pota, \kr, and \pam\ (from \ura\ decay chain), electron recoils from \ber, \pep, and \CNO\ solar neutrinos, and external $\gamma$-rays from \tleight, \bifo, and \pota.  We fixed the contribution from  \pp\ solar neutrinos to the SSM predicted rate (assuming MSW-LMA with $\tan^2\theta_{12}$=0.47$^{+0.05}_{-0.04}$, $\Delta m^2_{12}$=\powert{-5}{(7.6$\pm$0.2)}\,eV$^2$~\cite{bib:pdg2010}) and the contribution from \bor\ neutrinos to the rate from the measured flux~\cite{bib:sno}.  We fixed the rate of the radon daughter \pbfo\ using the measured rate of \Bipo\ delayed coincidence events.

Simultaneously to the fit of events surviving the TFC veto, we also fit the energy spectrum of events rejected by the veto, corresponding to the remaining 51.5\% of the exposure. We constrained the rate for every non-cosmogenic species to be the same in both data sets, since only cosmogenic isotopes are expected to be correlated with neutron production.

\begin{figure}[!t]
\centering
\includegraphics[width=\linewidth]{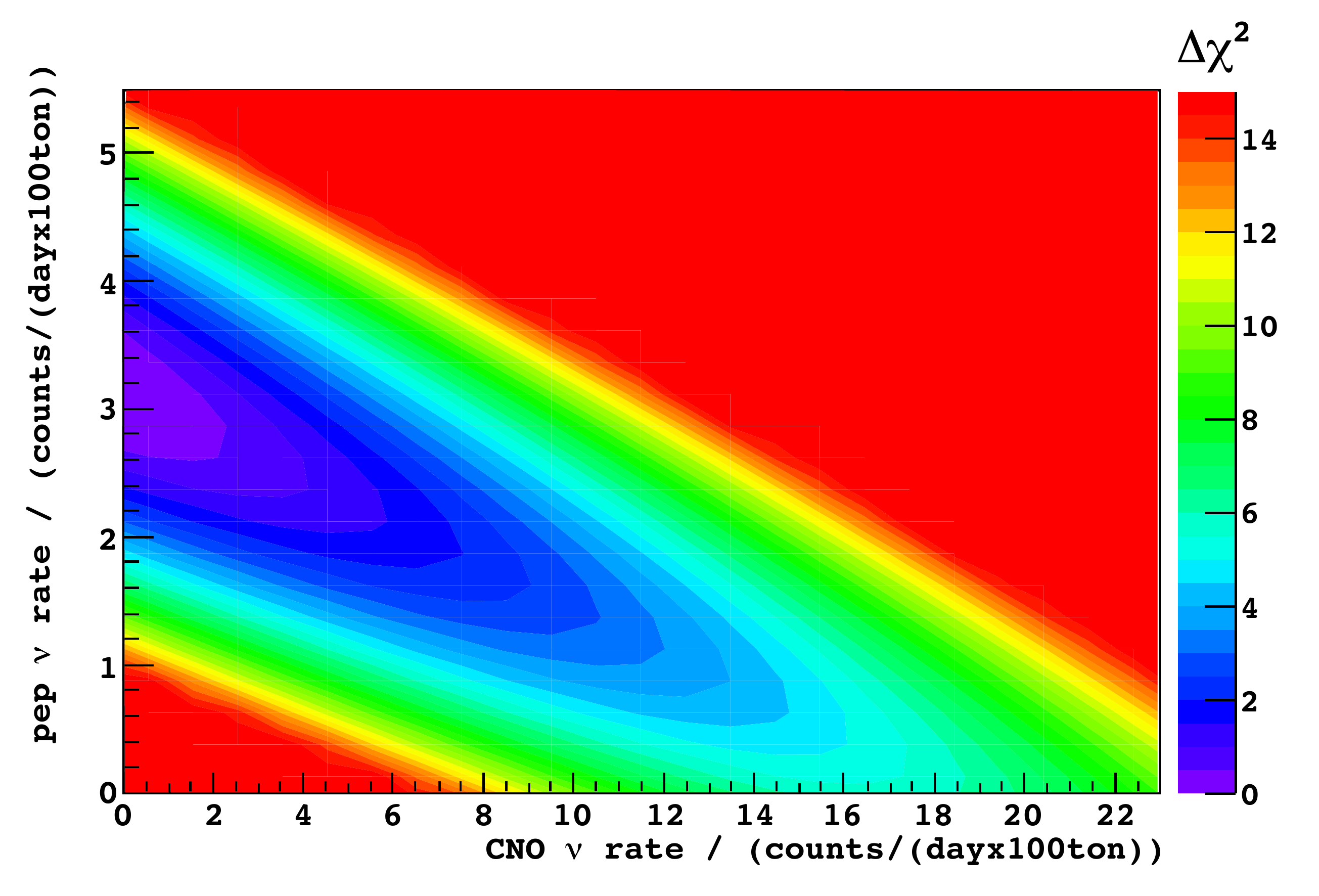}
\caption{
$\Delta\chi^2$ profile obtained from likelihood ratio tests between fit results where the \pep\ and \CNO\ neutrino interaction rates are fixed to particular values (all other species are left free) and the best-fit result.
}
\label{fig:dchi2}
\end{figure}

Fits to simulated event distributions, including all species and variables considered for the data fit, returned results for the \pep\ and \CNO\ neutrino interaction rates that were unbiased and uncertainties that were consistent with frequentist statistics.  These tests also yielded the distribution of best-fit likelihood values, from which we determined the p-value of our best-fit to the real data to be 0.3.  Table~\ref{tab:results-summary} summarizes the results for the \pep\ and \CNO\ neutrino interaction rates.  The absence of the solar neutrino signal was rejected at 99.97\%~C.L. using a likelihood ratio test between the result when the \pep\ and \CNO\ neutrino interaction rates were fixed to zero and the best-fit result.  Likewise, the absence of a \pep\ neutrino signal was rejected at 98\%~C.L.  Due to the similarity between the electron-recoil spectrum from CNO neutrinos and the spectral shape of \bite, whose rate is $\sim$10 times greater, we can only provide an upper limit on the \CNO\ neutrino interaction rate.  The 95\%~C.L. limit reported in Table~\ref{tab:results-summary} has been obtained from a likelihood ratio test with the \pep\ neutrino rate fixed to the SSM prediction~\cite{bib:ssm2011} under the assumption of MSW-LMA, (2.80$\pm$0.04)\,\cpd, which leads to the strongest test of the solar metallicity.  For reference, Fig.~\ref{fig:dchi2} shows the full $\Delta \chi^2$ profile for \pep\ and \CNO\ neutrino interaction rates.

\begin{table}[!t]
\begin{center}
\begin{tabular}{lccc}
\hline
\hline
$\nu$		&Interaction rate		&Solar-$\nu$ flux				&Data/SSM\\
			&[\cpd]				&[\powero{8}\flx]				&ratio\\
\hline
\pep			&$3.1 \pm 0.6_{\rm stat} \pm$ 0.3$_{\rm syst}$ &$1.6\pm0.3$				&$1.1\pm0.2$\\
\CNO		&$<7.9$ ($<7.1_{\rm stat\,only}$)				&$<7.7$					&$<1.5$\\
\hline
\hline
\end{tabular}
\end{center}
\caption{
The best estimates for the \pep\ and \CNO\ solar neutrino interaction rates.  For the results in the last two columns both statistical and systematic uncertainties are considered.  Total fluxes have been obtained assuming MSW-LMA and using the scattering cross-sections from \cite{bib:BahcallRadiativeCorrection, bib:pdg2010, bib:erlerRadCorr} and a scintillator $e^-$ density of \powert{29}{(3.307$\pm$0.003)}\,ton$^{-1}$.  The last column gives the ratio between our measurement and the High Metallicity (GS98) SSM~\cite{bib:ssm2011}.
}
\label{tab:results-summary}
\end{table}
\begin{table}[!t]
\begin{center}
\begin{tabular}{lcc}
\hline
\hline
Background	&Interaction rate		&Expected rate \\
			&[\cpd]				&[\cpd] \\
\hline
\kr			& $19^{+5}_{-3}$		&$30\pm6$~\cite{bib:bxbe7} \\
\bite			& $55^{+3}_{-5}$		&-- \\
\cele			& $27.4\pm0.3$		&$28\pm5$ \\
\cten			& $0.6\pm0.2$			&$0.54\pm0.04$ \\
\hesix		& $<2$				&$0.31\pm0.04$ \\
\pota			& $<0.4$				&-- \\
\pam			& $<0.5$				&$0.57\pm0.05$ \\
Ext. $\gamma$	& $2.5\pm0.2$			&-- \\
\hline
\hline
\end{tabular}
\end{center}
\caption{
The best estimates for the total rates of the background species included in the fit. The statistical and systematic uncertainties were added in quadrature.  The expected rates for the cosmogenic isotopes \cele, \cten\ and \hesix\ have been obtained following the methodology outlined in \cite{bib:bxb8}.  The expected \pam\ rate was determined from the \Bipo\ measured coincidence rate, under the assumption of secular equilibrium. Ext.~$\gamma$ includes the estimated contributions from \tleight, \bifo\ and \pota\ external $\gamma$-rays.
}
\label{tab:bkg}
\end{table}

The estimated \ber\ neutrino interaction rate is consistent with our measurement~\cite{bib:bxbe7}.  Table~\ref{tab:bkg} summarizes the estimates for the rates of the other background species.  The higher rate of \bite\ compared to \cite{bib:bxbe7} is due to the exclusion of data from 2007, when the observed decay rate of \bite\ in the FV was smallest.

Table~\ref{tab:syst} shows the relevant sources of systematic uncertainty.  To evaluate the uncertainty associated with the fit methods we have performed fits changing the binning of the energy spectra, the fit range and the energy bins for which the radial and pulse-shape parameter distributions were fit.  This has been done for energy spectra constructed from both the number of PMTs hit and the total collected charge in the event.  Further systematic checks have been carried out regarding the stability of the fit over different exposure periods, the spectral shape of the external $\gamma$-ray background and electron recoils from \CNO\ neutrinos, the fixing of \pbfo\ and \pp\ and \bor\ neutrinos to their expected values, and the exclusion of minor radioactive backgrounds (short-lived cosmogenics and decays from the \tho\ chain) from the fit.

\begin{table}[!t]
\begin{center}
\begin{tabular}{lc}
\hline
\hline
Source								&[\%] \\
\hline
Fiducial exposure						&$^{+0.6}_{-1.1}$ \\
Energy response						&$\pm4.1$ \\
\bite\ spectral shape						&$^{+1.0}_{-5.0}$ \\
Fit methods							&$\pm5.7$ \\
Inclusion of independent \kr\ estimate		&$^{+3.9}_{-0.0}$ \\
$\gamma$-rays in pulse shape distributions	&$\pm2.7$ \\
Statistical uncertainties in pulse shape distributions &$\pm5$ \\
\hline 
Total systematic uncertainty				&$\pm10$ \\
\hline
\hline
\end{tabular}
\end{center}
\caption{
Relevant sources of systematic uncertainty and their contribution in the measured \pep\ neutrino interaction rate. These systematics increase the upper limit in the \CNO\ neutrino interaction rate by 0.8\,\cpd.
}
\label{tab:syst}
\end{table}

\begin{figure}[!t]
\centering
\includegraphics[width=\linewidth]{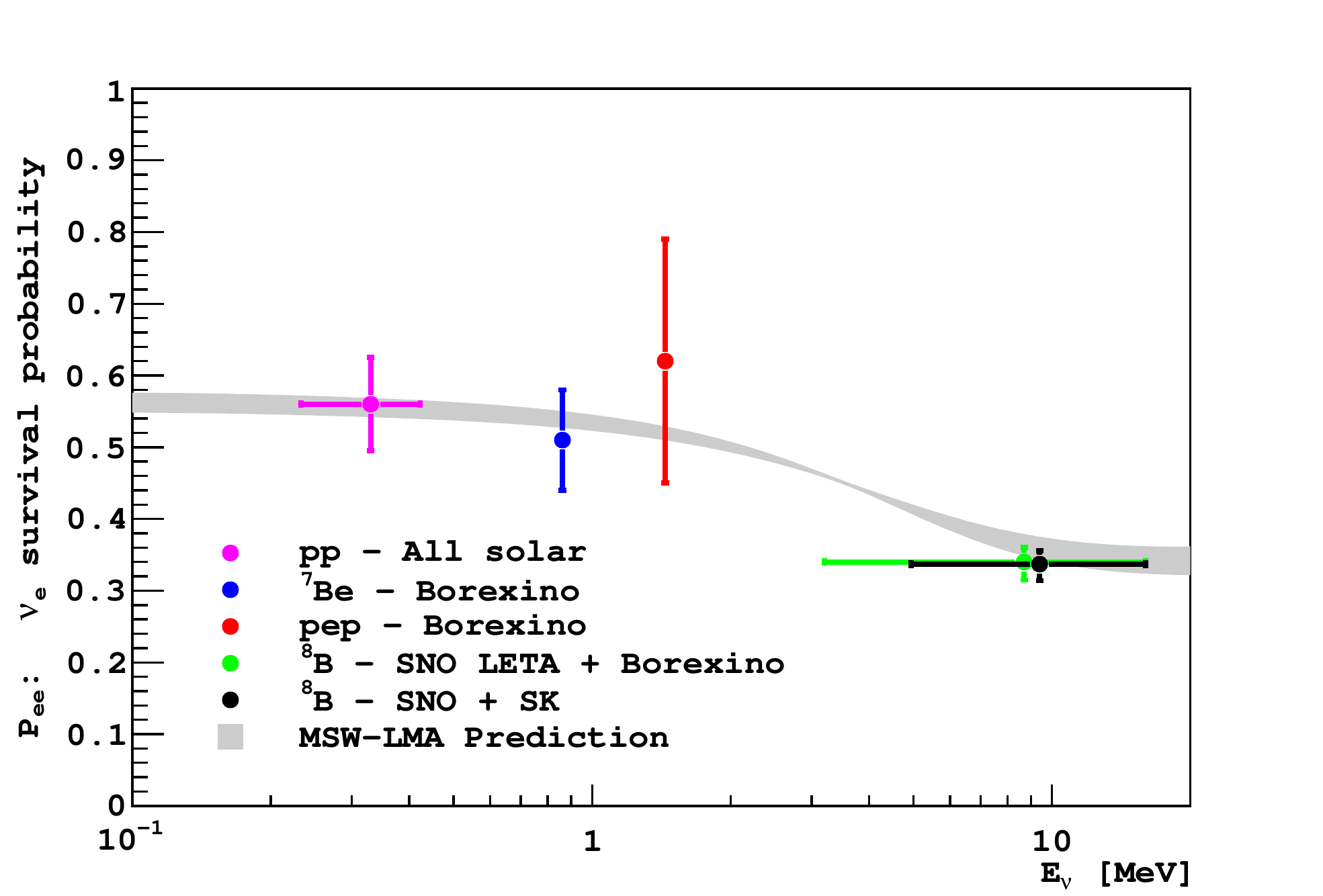}
\caption{
Electron neutrino survival probability as a function of energy.  The red line corresponds to the measurement presented in this letter.  The \pp\ and \ber\ measurements of \Pee\ given in~\cite{bib:bxbe7} are also shown.  The \bor\ measurements of \Pee\ were obtained from~\cite{bib:kamiokande,bib:sno,bib:bxb8}, as indicated in the legend.  The MSW-LMA prediction band is the $1\sigma$ range of the mixing parameters given in \cite{bib:pdg2010}.
}
\label{fig:pee}
\end{figure}

Table~\ref{tab:results-summary} also shows the solar neutrino fluxes inferred from our best estimates of the \pep\ and \CNO\ neutrino interaction rates, assuming the MSW-LMA solution, and the ratio of these values to the High Metallicity (GS98) SSM predictions \cite{bib:ssm2011}.  Both results are consistent with the predicted High and Low Metallicity SSM fluxes assuming MSW-LMA.  Under the assumption of no neutrino flavor oscillations, we would expect a \pep\ neutrino interaction rate in Borexino of (4.47$\pm$0.05)\,\cpd; the observed interaction rate disfavors this hypothesis at 97\%~C.L.  If this discrepancy is due to $\nu_e$ oscillation to $\nu_\mu$ or $\nu_\tau$, we find \Pee=0.62$\pm$0.17 at 1.44\,MeV. This result is shown alongside other solar neutrino \Pee\ measurements in Fig.~\ref{fig:pee}. The MSW-LMA prediction is shown for comparison.

We have achieved the necessary sensitivity to provide, for the first time, evidence of the rare signal from \pep\ neutrinos and to place the strongest constraint on the \CNO\ neutrino flux to date.  This has been made possible by the combination of the extremely low levels of intrinsic background in Borexino, and the implementation of novel background discrimination techniques.  This result raises the prospect for higher precision measurements of \pep\ and \CNO\ neutrino interaction rates, if the next dominant background, \bite, is further reduced by scintillator re-purification.

The Borexino program is made possible by funding from INFN (Italy), NSF (USA), BMBF, DFG and MPG (Germany), NRC Kurchatov Institute (Russia), and MNiSW (Poland). We acknowledge the generous support of the Gran Sasso National Laboratories (LNGS).

\end{document}